\documentclass[aps,prl,twocolumn,superscriptaddress,showpacs,showkeys]{revtex4}

\usepackage{graphicx}% Include figure files
\usepackage{amsmath}

\begin{document}

%\preprint{}

\title{Polarization effects in diffraction of light on a planar chiral structure}


\author{S. L. Prosvirnin}
%\email[]{prosvirn@rian.kharkov.ua}
%\thanks{}
\affiliation{Institute of Radio Astronomy, National Academy of
Sciences of Ukraine, Kharkov, 61002, Ukraine}

\author{N. I. Zheludev}
\email{n.i.zheludev@soton.ac.uk}
\homepage{www.nanophotonics.phys.soton.ac.uk}
%\thanks{}
\affiliation{School of Physics and Astronomy, University of
Southampton, SO17 1BJ, UK}


\date{\today}

\begin{abstract}
We analyzed polarization changes of light diffracted on planar
chiral array from the standpoint of the Lorentz reciprocity lemma
and found the bi-orthogonality in the polarization eigenstates for
the waves diffracting though the grating in the opposite
direction. Both reciprocal and non-reciprocal component in the
polarization azimuth rotation of the diffracted light were
identified. The structural chirality of the array arrangement and
the chirality of individual elements of the array give rise to
polarization effects.
\end{abstract}
\maketitle
%\section{}
%\subsection{}
%\subsubsection{}
Recently we reported that planar (2D) chiral structures affect the
polarization state of light in an enantiomeric fashion, similarly
to three--dimensional chiral media \cite{zheludev1}. However
polarization phenomena on diffraction from planar chiral
structures have never been studied theoretically before, leaving
the fundamental properties of 2D chirality not fully understood.
Here we report on the results of a theoretical investigation of
polarization changes for light diffracted on regular arrays of
planar chiral metallic structures from the standpoint of the
Lorentz reciprocity theorem.
%By analyzing propagation of light in
%two opposite directions we found that both reciprocal and
%non-reciprocal component in the polarization azimuth rotation of
%the diffracted light.
By analyzing propagation of light in two opposite directions we
have identified a strong component in polarization effect on
diffraction that can be induced either by the chirality of the
individual elements of the array, or by arranging non-chiral
elements of an array in a chiral fashion.

\begin{figure}
 \includegraphics{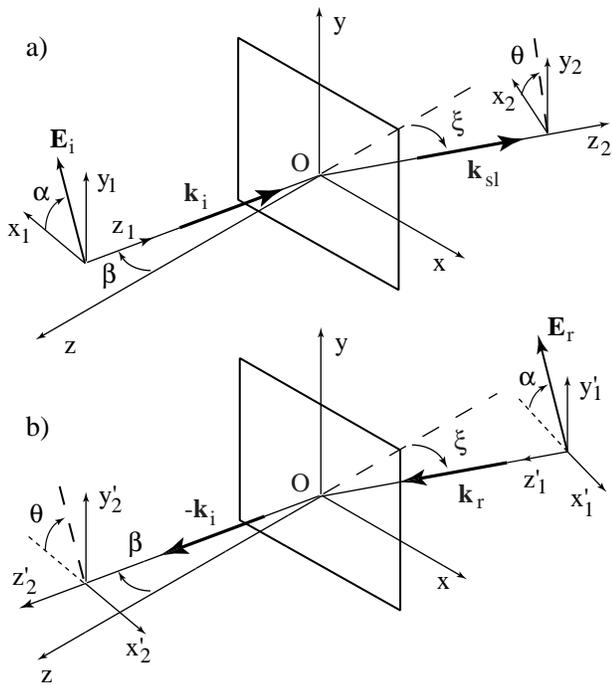}
 \caption{Coordinate systems and waves in the "direct" (a) and
 "reversed" (b) diffraction scenarios. \label{coord}}
 \end{figure}

\begin{figure}
 \includegraphics[width=6cm]{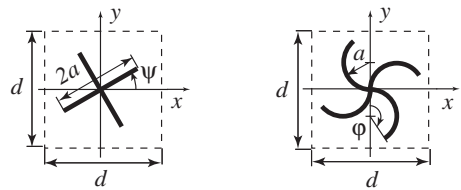}
 \caption{Structural elements of the arrays: planar straight cross tilted against
 the array greed on the tilt angle $\psi$, and
 chiral right-handed gammadion with bending angle $\varphi$. \label{II&SS}}
 \end{figure}

Let us consider a planar square periodic array of metallic
elements of thickness $t$ with equal pitch $d$ along the axes $x$
and $y$ placed between planes $z=0$ and $z=-t$ (see
Fig.~\ref{coord} and Fig.~\ref{II&SS}). If a plane electromagnetic
wave
\begin{equation}\label{E1i}
 {\mathbf{E}}_{i}={\mathbf A}_i e^{-i{\mathbf{k}}_{i}{\mathbf r}}
\end{equation}
of unit amplitude and polarization vector ${{\mathbf A}_i}$ is
incident on the array from the region corresponding to $z>0$, the
transmitted field may be written as a summation over all
diffracted waves, numbered by integer indices $q$ and $p$
\begin{equation}\label{E1}
  {\mathbf{E}}_{t}=
      \sum_{q,p=-\infty}^\infty {\mathbf{a}}_{qp}
    e^{ -i{\mathbf{k}}_{qp}({\mathbf r}+{\mathbf e}_z t)},
    \quad z<-t
\end{equation}
where $\mathbf{a}_{qp}$ and $\mathbf{k}_{qp}$ are amplitudes and
wave-vectors partial of diffracted waves and
\begin{eqnarray*}
{\mathbf{k}}_{qp} = {\mathbf{g}}+{\mathbf{h}}_{qp}-{\mathbf
e}_z \sqrt{k^2-|{\mathbf{g}}+{\mathbf{h}}_{qp}|^2},\\
{\mathbf{h}}_{qp} = 2\pi(q {\mathbf e}_x + p {\mathbf e}_y)/d.
\end{eqnarray*}
Here ${\mathbf e}_x$, ${\mathbf e}_y$ and ${\mathbf e}_z$ are unit
vectors along the axes $x$, $y$ and $z$, $\mathbf{g}$ is the
component of ${\mathbf{k}}_{i}$ transverse to the axis $z$ and
$k=|\mathbf{k}_i|$. Let us now consider a "reversed" wave with
polarization vector ${{\mathbf A}_r}$ approaching the array from
the opposite side of the structure ($z<-t$) along the direction of
one of the partial diffracted waves of the "direct" scenario, with
indices $s$ and $l$ and the wave vector ${\mathbf{k}}_{r} =
-{\mathbf{k}}_{sl}$ ,
\begin{equation}\label{E2i}
{\mathbf{E}}_{r}={\mathbf A}_r e^{-i{\mathbf{k}}_{r}({\mathbf
r}+{\mathbf e}_z t)}.
\end{equation}
In the region $z>0$ this wave will produce diffracted waves  with
amplitudes ${\mathbf{b}}_{qp}$ . This corresponds to the
"reversed" scenario of diffraction. The Lorentz reciprocity lemma
\cite{kong} applied to the field superposition in the volume
bounded by surface $S$ that consists of planes $x=\pm d/2$, $y=\pm
d/2$, $z=z_1>0$, and $z=z_2<-t$ may be written in the following
form:
\begin{equation}\label{Lor}
  \oint_S \left\{ [\tilde{\mathbf{E}}_i \times \tilde{\mathbf{H}}_r]
  -[\tilde{\mathbf{E}}_r \times \tilde{\mathbf{H}}_i] \right\}d{\boldsymbol{\sigma}}=0
\end{equation}
where $\tilde{\mathbf{E}}_i$, $\tilde{\mathbf{H}}_i$ and
$\tilde{\mathbf{E}}_r$, $\tilde{\mathbf{H}}_r$ are electric and
magnetic fields created by waves incident from opposite
directions. By using corresponding field expressions it may be
shown from formula (4) that
\begin{equation}\label{bb}
\sqrt{k^2-|{\mathbf{g}}|^2} ({\mathbf A}_i \cdot {\mathbf
b}_{sl})=\sqrt{k^2-|{\mathbf{g}}+{\mathbf{h}}_{sl}|^2} ({\mathbf
A}_r \cdot {\mathbf a}_{sl}).
\end{equation}
The equality (\ref{bb}) constitutes the universal relation between
the amplitudes of partial waves in the "direct" and "reversed"
diffraction scenarios.
% if the process complies with the Lorentz reciprocity [\ref{Lor}].
Scattering processes are often described
in terms of $2\times 2$ transformation matrices, relating
Cartesian components of electric fields in coordinate frames of
incident and scattered waves. For the "direct" ($\hat D$) and
"reversed" ($\hat R$) scenarios these matrices for the incident
and partial diffracted waves can be introduced as follows:
${\mathbf a}_{sl}=\hat D {\mathbf A}_i,$ ${\mathbf b}_{sl}=\hat R
{\mathbf A}_r.$ It may be shown from equation (5) that these
matrices are linearly related and mutually transposed:
\begin{equation}\label{tt}
R_{nm} = c (2\delta_{mn} -1)D_{mn},
\end{equation}
where $\delta_{mn}$ is Kroneker index and
$c=\sqrt{k^2-|\mathbf{g}+{\mathbf{h}}_{sl}|^2}/
\sqrt{k^2-|\mathbf{g}|^2}$. For the purpose of analyzing the
polarization eigenstates of the diffraction process it is
instructive to present both scattering matrices in the coordinate
frame of the direct scenario (see Fig.~1,a) where the operator of
the reversed scattering process acts on the complex-conjugated
field amplitudes. Here the polarization eigenstates are simply two
linearly independent eigenvectors of matrices $\hat R^\ast$ and
$\hat D$. The relation between them may be derived from equation
(\ref{tt}), which, when converted to the coordinate frame of the
direct scenario, gives $R_{nm}^\ast = c D_{mn}^\ast$. It follows
from the theory of matrix operators that eigenvectors of the
Hermitian-conjugated matrices  with elements $D_{nm}$ and
$D_{mn}^\ast$, and therefore of matrices $\hat R^\ast$ and $\hat
D$, are biorthogonal or, in terms of polarization eigenstates, are
represented by antipode points on the Poincare Sphere, as shown on
Fig.~3. In general, the point representing the first eigenstate in
the direct scenario $1d$ is an antipode to one of the points,
which represents the eigenstates of the reversed scenario (this
point is designed as  $2r$ in Fig.~3). However eigenstate $1d$
does not necessary coincide with eigensate $1r$ of the reversed
scenario, nor eigenstate $2d$ coincide with $2r$. Therefore, the
polarization eigenstates in the direct and reversed scenarios
presented in the coordinate frame of the direct scenario could be
different. Such a situation takes place if the complex matrix
$\hat D$ is an asymmetric or even a non-diagonal matrix.

We found that scattering matrices of non-zero order diffraction on
periodic plano-chiral arrays, where chirality is due to either
structural chirality, or chirality of individual elements of the
array, are either asymmetric or non-diagonal. Below we will
illustrate these properties by numerical modelling of the
diffraction process for various planar chiral gratings. We
calculated the fields and polarization characteristics of light
diffracted on gratings numerically using the method described in
reference \cite{prosv}. It is based on a vector integral equation
for the surface current induced by the light wave on the array
particles. The equation is derived with boundary conditions for
ideal metallic structures that assume a zero value for the
tangential component of the electric filed on the metal. The
integral equation is then reduced to an algebraic equation set by
use of the Galerkin technique.

\begin{table*}%[H] add [H] placement to break table across pages
 \caption{Polarization Eigenstates (PES) for various diffraction processes presented in the "direct" scenario coordinate frame (all angles are measured in degrees, subscripts $d$ and $r$ denote "direct and "reversed" scenarios) \label{t1}}
 \begin{ruledtabular}
 \begin{tabular}{|c|c|c|c|c|c|}
  &\multicolumn{2}{c|}{"Direct" Scenario}&\multicolumn{2}{c|}{"Reversed" Scenario}& \\
           \cline{2-5}
Structure & 1-st PES & 2-nd PES & 1-st PES & 2-nd
 PES & Type of diffraction\\

 & (degrees) & (degrees) & (degrees) & (degrees) & \\
            \hline
Straight crosses   & $\theta_{1d}=0.00$ & $\theta_{2d}=90.00$ & $\theta_{1r}=0.00$ & $\theta_{2r}=90.00$ & No chiral effect\\
$\psi=0$; $\beta=0$  & $\eta_{1d}=0.00$ & $\eta_{2d}=0.00$ & $\eta_{1r}=0.00$ & $\eta_{2r}=0.00$ & $\theta_{1d}=\theta_{1r}$, $ \eta_{1d}=\eta_{1r}$\\
            \hline
Straight crosses   & $\theta_{1d}=-14.4$ & $\theta_{2d}=77.1$ & $\theta_{1r}=-12.9$ & $\theta_{2r}=75.6$ &  Chiral effect is present\\
$\psi=+15$; $\beta=0$ & $\eta_{1d}=-0.05$ & $\eta_{2d}=0.08$ & $\eta_{1r}=-0.08$ & $\eta_{2r}=0.05$ &  $\theta_{1d}-\theta_{1r}=-1.5$, $\theta_{2d}-\theta_{2r}=1.5 $\\
            \hline
Straight crosses   & $\theta_{1d}=14.4$ & $\theta_{2d}=-77.1$ & $\theta_{1r}=12.9$ & $\theta_{2r}=-75.6$ &  Chiral effect is present\\
$\psi=-15$; $\beta=0$ & $\eta_{1d}=0.05$ & $\eta_{2d}=-0.08$ & $\eta_{1r}=0.08$ & $\eta_{2r}=-0.05$ &  $\theta_{1d}-\theta_{1r}=1.5$, $\theta_{2d}-\theta_{2r}=-1.5 $\\
            \hline
Right gammadions   & $\theta_{1d}=6.1$ & $\theta_{2d}=-26.6$ & $\theta_{1r}=63.5$ & $\theta_{2r}=-83.9$ &  Chiral effect is present\\
$\varphi=120$; $\beta=0$ & $\eta_{1d}=4.38$ & $\eta_{2d}=-0.02$ & $\eta_{1r}=0.12$ & $\eta_{2r}=-4.42$ &  $\theta_{1d}-\theta_{1r}=-57.4$, $\theta_{2d}-\theta_{2r}=57.3 $\\
            \hline
Left gammadions   & $\theta_{1d}=-6.1$ & $\theta_{2d}=26.6$ & $\theta_{1r}=-63.5$ & $\theta_{2r}=83.9$ &  Chiral effect is present\\
$\varphi=120$; $\beta=0$ & $\eta_{1d}=-4.38$ & $\eta_{2d}=0.02$ & $\eta_{1r}=-0.12$ & $\eta_{2r}=4.42$ &  $\theta_{1d}-\theta_{1r}=57.4$, $\theta_{2d}-\theta_{2r}=-57.3 $\\
            \hline
Right gammadions   & $\theta_{1d}=9.9$ & $\theta_{2d}=-80.0$ & $\theta_{1r}=9.9$ & $\theta_{2r}=-80.0$ &  Chiral effect is present\\
$\varphi=120$; $\beta=\beta_2$ & $\eta_{1d}=5.2$ & $\eta_{2d}=5.2$ & $\eta_{1r}=-5.2$ & $\eta_{2r}=-5.2$ &  $\theta_{1d}=\theta_{1r}$, $\eta_{1d}=-\eta_{1r}$\\
            \hline
Left gammadions   & $\theta_{1d}=-9.9$ & $\theta_{2d}=80.0$ & $\theta_{1r}=-9.9$ & $\theta_{2r}=80.0$ &  Chiral effect is present\\
$\varphi=120$; $\beta=\beta_2$ & $\eta_{1d}=-5.2$ & $\eta_{2d}=-5.2$ & $\eta_{1r}=5.2$ & $\eta_{2r}=5.2$ &  $\theta_{1d}=\theta_{1r}$, $\eta_{1d}=-\eta_{1r}$\\
% Lines of table here ending with \\
 \end{tabular}
 \end{ruledtabular}
 \end{table*}

In our modelling we concentrated on planar chiral arrays of 442
symmetry wallpaper group and calculated the polarization
eigenstates of the diffraction process and polarization changes
occurring in the diffracted wave for different incident
polarizations.
 \begin{figure}
 \includegraphics[width=6cm]{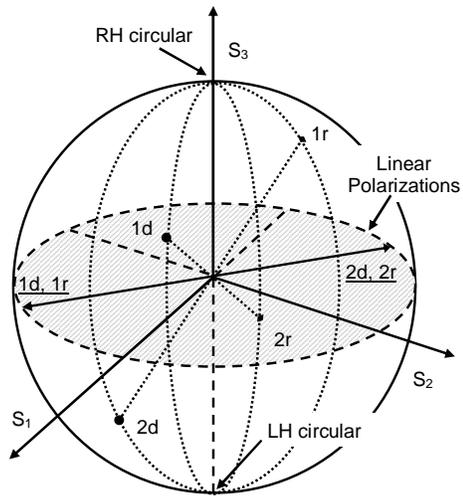}
 \caption{Schematic representation of diffraction on the Poincar\'{e}
 sphere. For a chiral grating polarization eigenstates in the "direct"
 scenario ($1d$ and $2d$) and "reversed" scenario ($1r$ and $2r$)
 are elliptical. Polarization eigenstates (underlined) for non-chiral
 gratings are mutually perpendicular linear polarization.}
 \label{sphere}
 \end{figure}

\begin{figure}
 \includegraphics{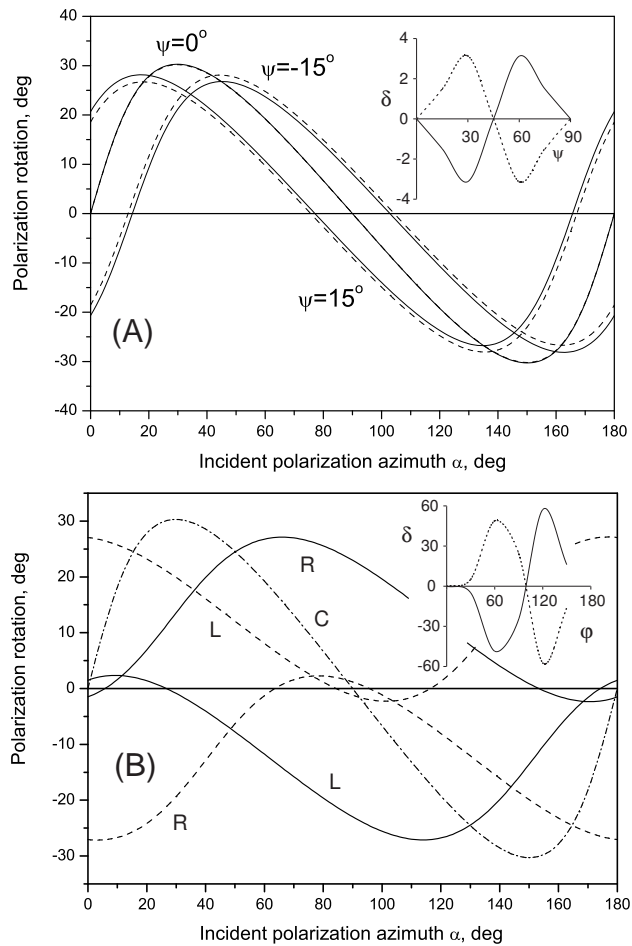}
 \caption{
Polarization azimuth rotation $\Delta=\theta-\alpha$ on
diffraction from chiral arrays as a function of incident
polarization azimuth, straight line - "direct" scenarios, dashed
line - "reversed" scenarios (all results are presented in the
direct scenario coordinate frame). The influence of chirality is
manifested as a split between corresponding solid and dashed
lines: A) Array of straight crosses. The insert shows the chiral
difference in polarization azimuth for the two polarization
eigenstates as a function of the tilt angle of the crosses; B)
Array of left (L) and right (R) gammadions.  The inset shows the
chiral difference in polarization azimuth for the two polarization
eigenstates as a function of gammadion bending angle. \label{rII}}
 \end{figure}

We studied diffraction for two different incident angles $\beta$,
at $\beta_1=0$ and at $\beta_2=\arcsin(\pi /kd)$. In the first
case, the diffracted wave ($q=1$, $p=0$) propagates at angle
$\xi=\xi_1=\arcsin(2\pi/kd)$ to the array. In the second case the
diffracted wave ($q=-1$, $p=0$) the same angle makes to the array
as the incident wave $\xi=\xi_2=-\beta_2$ (for definitions of
angles see Fig.1). The wave's polarization azimuth $\theta$ and
degree of ellipticity $\eta$ were calculated from the Cartesian
field amplitudes using the standard definitions: $\tan
2\theta=s_2/s_1$, $\sin 2\eta=s_3/s_0$, where $s_i$ are the Stokes
parameters. The results of our analysis for $d=4\;\mu\text{m}$,
$\lambda=2\pi/k=1520$ nm, $\beta_1=0$, $\xi_1=22.3^\circ$,
$\beta_2=11.0^\circ$, $\xi_2=-11.0^\circ$ are summarized in
Table~\ref{t1} and Fig.~\ref{rII}. We considered an array without
a substrate. The width of the metal strips was equal to 0.05
$\mu$m.

In optics polarization elements are often classified as reciprocal
or non-reciprocal depending on whether their effect on the
polarization state of the transmitted light is the same or
different for light propagating in the opposite directions. This
understanding of optical reciprocity which we will use below is
somewhat different from the general, more tolerant definition of
reciprocity based on the Lorentz lemma. For the purpose of
comparison of the polarization transformations for opposite
directions of light propagation, in the table and figures the
polarization parameter of the waves are converted into the
coordinate frame of the direct scenario. In such a presentation if
the values of polarization azimuth rotation in the direct and
reversed scenarios are the same, the rotation is truly
non-reciprocal. On contrary, a difference between the values of
polarization azimuth rotation in the direct and reversed scenarios
would represent a reciprocal component of the polarization change,
that is analogous to the optical activity effect in a chiral
liquid.

The calculations revealed that:

i) For all diffraction processes involving twisted or non-twisted
arrays, equalities (6) are held to within the numerical accuracy
of the method. They are thus compatible with the Lorentz lemma.

ii) For the arrays of straight crosses polarization azimuth
rotation in opposite directions have opposite signs due to the
difference in the efficiency of diffraction for perpendicular
polarization components (line C on Fig.~4(B)). This
non-reciprocity of polarization azimuth rotation is analogous to
the polarization rotation non-reciprocity in dichroic media due to
anizotropic dissipation.

iii) No polarization rotation is seen in the non-diffracted part
of the beam at the normal incidence. It's polarization eigenstates
are the same in both directions and for any type of array.

iv) From Table~\textrm{I}, one can see that for non-zero order
asymmetrical diffraction ($\mid\xi\mid\neq\mid\beta\mid$) when
individual structural elements of the array are twisted rosettes
polarization azimuths of eigenstates for the "direct" and
"reversed" scenarios are resolutely different. The difference
between the polarization azimuths of the eigensatates depends on
the rosette curvature angle $\varphi$ and reaches a maximum of
about $57^\circ$ at $\varphi=120^\circ$. The difference in the
eigenstates vanishes at rosette bending angle $\varphi=95^\circ$.

v) When individual structural elements of the array are twisted
rosettes, polarization azimuth rotation on diffraction has both
reciprocal and nonreciprocal components. The non-reciprocal
component of the polarization azimuth rotation is due to a
difference in the efficiency of diffraction for perpendicular
polarization components. The corresponding oscillating dependence
of the nonreciprocal rotation on the incident angle is shifted in
respect of line C corresponding to straight crosses. It is shifted
along the incident polarization azimuth axis towards left for left
rosettes and towards right for right rosettes. The split between
corresponding solid and dashed lines in Fig.~4 indicates the
reciprocal component of the polarization azimuth rotation
analogous to optical activity.

vi) Nonreciprocity of polarization rotation in the diffraction
process is evident when a diffracted light wave is reflected
straight back towards the twisted planar structure by a mirror,
and then diffracts again. The polarization state of the returning
light after the second diffraction is different from that of the
incident light, even if the incident light was an eigenstate in
the forward direction. For an array of rosettes with
$\varphi=120^\circ$, the two incident eigenstates and
corresponding returning polarizations have azimuths different by
$27^\circ$ and $21^\circ$.

Therefore, polarization effects on diffraction from planar chiral
grating can be induced by either structural chirality or the
chirality of individual elements of the array. However, no
polarization rotation compatible with the Lorentz lemma is
possible for a wave transmitting through or reflected from a
planar chiral structure at normal incidence, as the scattering
matrices are diagonal in this case.

Finally, we shall note that our analysis is underpinned by the
Lorentz lemma while our computational method is compatible with
it. It shall be noted, however, that  the recent observation of
broken parity and time reversal evident in polarized optical
images of planar chiral structures \cite{zheludev2} calls for
re-examining the validity of the Lorentz lemma for planar chiral
structure.

\begin{acknowledgments}
The authors thank A.Papakostas, A.Potts, D.Bagnall and K.MacDonald
for fruitful discussions and acknowledge the support of the
Science and Engineering Research Council (UK).
\end{acknowledgments}

\bibliography{PolStates}

\end{document}